\documentclass[german,psfrag,titlepage,epsf,amsmath,a4paper,15pt]{afi}
\usepackage{graphics}
\usepackage{epsfig}
\usepackage{amssymb}
\usepackage{url}
\makeindex
\begin{document}

\newcommand{\bfm}{\boldsymbol}
\def\partialt{\partial_t}
\def\bnab{\bnabla \! \!}
\def\bb{{\bf B}}
\def\simgeq{\mathrel{\smash{\mathop{\raise2pt\hbox{$>$}}\limits_%
   {\smash{\raise4pt\hbox{$\sim$}}}}\vphantom\geq}}

\def\@abstract[#1]{%
  \global\@hasabstracttrue
  \hyphenpenalty\sv@hyphenpenalty     % restore \hyphenpenalty
  \global\setbox\t@abstract=\vbox\bgroup
  \leftskip\z@
  \@rightskip\z@ \rightskip\@rightskip \parfillskip\@flushglue
   \@abstractsize                      % Text in 9/11
  \parindent 1em                      % \parindent in abstract
  \noindent {\bfseries\abstractname}  % caption `Abstract' (bold)
  \vskip 0.5\@bls    % half a line of space below
\noindent\ignorespaces
}

\pagenumbering{roman}

%\skippage
\pagenumbering{arabic}
\setcounter{page}{1}

\iupsize
\baselineskip 18pt

\chapter{\noindent
Particle astrophysics from the cold: Results and perspectives of IceCube
}
\vspace*{-0.5cm}
C. de los Heros\par
{\noindent (for the IceCube collaboration)$^*$}
\footnote{Prepared for the proceedings of the first AFI symposium, Innsbruck, 19-20/10/2007 . To be published by 
Innsbruck University Press. Eds S.~D.~Bass, F.~Schallhart and B.~Tasser}\par

\vspace{0.5cm}

\noindent{\normalsize \it
Department of Physics and Astronomy, Uppsala University, Uppsala, Sweden.}

\vspace{0.5cm}

\noindent{
\normalsize
 We discuss results of the AMANDA neutrino telescope, in operation at the South Pole since 2000, and 
present the status and scientific potential of its km$^3$ extension, IceCube.
}

\vspace{0.5cm}

\section{Introduction}
\label{sec:intro}

In 1931 Victor Hess founded a research station 2300m up at the Hafelekar mountain, not very far from the premises 
of this meeting, for observing and studying cosmic rays.  Hess received the Nobel Prize in Physics in 1936 
for the discovery of ``cosmic radiation'', and in his Nobel lecture he already identified the key to the development 
of the new opened field ``In order to make further progress, particularly in the field of cosmic rays, it will be 
necessary to apply all our resources and apparatus simultaneously and side-by-side''~\cite{Hess:36a}. Seventy years 
after his words, the field of astroparticle physics has come to maturity precisely from 
the close cooperation of optical and gamma ray telescopes, air shower arrays and neutrino telescopes. \par
 Acceleration of particles to the extreme energies detected today (up to 10$^{20}$ eV) is assumed to be driven by shock fronts propagating in hot and dense regions of ionized matter. Such conditions are expected to be found in the neighborhood of accreting 
objects, like AGN or micro-quasars, or in extreme explosions like  Gamma Ray Bursts (GRBs).   There are compelling 
theoretical arguments to expect neutrino production from these sites as well. Accelerated protons must interact with the 
ambient matter or radiation, producing secondary charged pions and kaons which decay to neutrinos. Neutral pions 
will also be produced and they will decay into $\gamma\gamma$, giving a normalization of the neutrino 
flux with the gamma ray flux. A neat example of the need for ``using apparatus simultaneously'' mentioned by Hess. \par
  In this proceedings we will present results from the AMANDA neutrino telescope on the searches for cosmic neutrinos and 
discuss the status and first physics results from its km$^3$ successor, IceCube. 
Neutrino telescopes are not only astrophysical instruments, 
but can be used to address topics in cosmology and particle physics. We will also report on the AMANDA results on searches 
for WIMPs, monopoles and non-standard oscillation scenarios.

\section{The AMANDA and IceCube detectors}
\label{sec:detector}
 As of January 2008, the IceCube Neutrino Observatory~\cite{IceCube:Performance} is close to half size of  
its final design and it is already the largest neutrino telescope in the world. 
 The observatory consists of a 1 km$^3$ ice array, IceCube, and a surface air-shower 
array, IceTop. The ice array will consist of up to 80 strings 
holding 60 digital optical modules (DOMs) each, deployed at depths between 1450~m - 2450~m 
near the geographic South Pole. The DOMs are vertically separated by 17 m, 
while the strings are arranged in a triangular grid with an inter-string separation 
of 125 m.  Each IceCube DOM contains a 25-cm Hamamatsu photomultiplier tube with electronics 
for in-situ digitization and timing of the photomultiplier waveforms, as well as a LED flasher 
board for calibration purposes. The observed dark noise rate of the DOMs is about 700 Hz. \par
 IceCube is designed to detect the Cherenkov radiation of secondaries  produced in 
neutrino interactions. For the $\nu_{\mu}$ channel, the 
Earth is used as a filter and only up-going muon tracks are considered, due to the overwhelming 
down-going atmospheric muon background. The detector monitors therefore the northern sky. 
For the $\nu_{e}$ and $\nu_{\tau}$ channels, the signature is the particle cascade 
produced in the neutrino interactions close or inside the detector, and the whole sky can then 
be monitored. \par
 The IceTop surface array uses the same DOM architecture as IceCube. The array will consist of 
80 stations, one near the surface location of each IceCube string. A station consists of two
  ice tanks, with two DOMs in each one operated at different gains  to 
increase the tank dynamic range. IceTop will detect the charged particle component of air showers above 10$^{14}$ eV. \par
 Historically, the first neutrino detector at the South Pole was AMANDA. 
Completed in 2000, the detector consists of 677 optical modules (20-cm photomultiplier 
tubes housed in glass spheres), deployed in 19 strings at depths between 1450~m and
 2000~m. The strings are arranged in three approximately concentric circles of 40~m, 100~m and 200~m in 
diameter. The AMANDA optical modules are simpler than the IceCube DOMs, not containing 
any embedded digitizing hardware. Signal processing and triggering is done at the surface. 
%Originally, the AMANDA DAQ consisted of peak-sensing ADCs, which did not provide any waveform information. 
%Beginning in 2003, the AMANDA DAQ was upgraded to include full waveform readout, increasing the dynamic range of the OMs 
%and providing a similar information content to the IceCube DOMs. Indeed AMANDA was fully 
AMANDA was fully incorporated into the IceCube detector as a subsystem in 2007 and it is now part of the IceCube trigger 
system. With its denser string spacing (typically 30 m) AMANDA can be used as a low energy array. IceCube strings 
surrounding AMANDA can be used as a veto to define contained events, lowering the energy 
threshold to a few tens of GeV.

\section{Results from AMANDA}\label{sec:AMA_results}
\subsection{The Galactic plane}
 Cosmic rays interacting with the interstellar medium in the galaxy are a 
source of guaranteed neutrinos, produced through secondary pion and kaon
decays. From the geographical location of AMANDA, only the outer region of 
the galactic plane, between longitude 33$^\circ<\delta<$ 213$^\circ$, lies below the horizon 
and can therefore be monitored. We have searched the data collected between 2000 and 2003, 3329 neutrino events, 
 for a possible enhancement of the neutrino flux from the galactic disk. 
 Assuming a Gaussian shape of the distribution of matter in the 
galactic disk, with a width of 2.1$^\circ$, and a $E_{\nu}^{-2.7}$
spectrum, we obtain a 90\% CL limit on the neutrino flux of 4.8$\times 10^{-4}$ GeV$^{-1}$
cm$^{-2}$sr$^{-1}$ s$^{-1}$ in the range 0.2 TeV $<E_{\nu}<$ 40 TeV~\cite{AMANDA:galactic_plane}.

\subsection{Searches for a cosmic neutrino flux}

\noindent{\bf a) Steady Point sources:} 
 The search for point sources of neutrinos is performed by looking for statistical excesses 
of events in narrow angular regions in the sky, determined by the angular resolution of the detector (about 
2$^\circ$ for this analysis). The search can be done in a generic way, looking for ``hot spots'' with 
respect to the average background, or by looking at the position of predefined candidate objects. 
In the latter case, the background is estimated from the data off-source, in the same declination band 
as the candidate object. These searches are done exclusively for muon neutrinos due to the better pointing 
resolution, and therefore are restricted to the northern sky. 
However the search is also sensitive to tau neutrinos through the muon produced in the tau decay,  
so the upper limits in table~1.1 refer to the combined flux $\mathrm{\Phi_{\nu_\mu} + \Phi_{\nu_\tau}}$. 
See~\cite{AMANDA:5yr_pointsource} for details.\par
  The results mentioned here were obtained with the combined
 data sets of the years 2000-2004. It amounts to a total of 1001 days of 
live-time, and the sample contains 4282 upward going neutrino events. 
Figure~\ref{fig:skyplot} shows the significance map of the northern  
sky in galactic coordinates. The map is compatible with a random 
distribution of sources, the hottest spot having a 92\% probability of 
being a random fluctuation. \par
  The same data set has been used to search for a neutrino flux from the direction 
of known objects. For this search we have used 32 sources known to be gamma and/or
X-ray emitters, like Blazars, micro-quasars or supernova remnants.  Table~1.1 shows the neutrino flux 
limits obtained for a few selected sources assuming an $E^{-2}$ neutrino energy spectrum~\cite{AMANDA:5yr_pointsource}.\par

\begin{figure}[t]
\epsfxsize=4.0in
\centerline{\epsffile{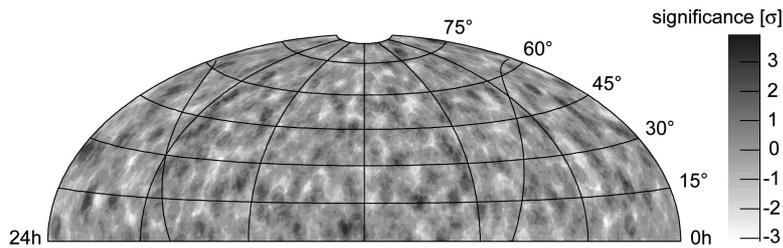}}
\caption[*]{
Significance sky map using 5 years of AMANDA data. The figure shows the deviation 
 from a uniform background. 
\label{fig:skyplot}
}
\end{figure}

\begin{table}[t]
\label{table:objects}
 \begin{center}
% \begin{footnotesize}
  \begin{tabular}{lcccccc|lcccccc}\hline\hline
      Candidate & $\delta$($^\circ$) & $\alpha$(h)    &
                $n_{\mathrm{obs}}$ & $n_{b}$          &&
                $\Phi_{\nu}^{\mathrm{lim}}$           &
      Candidate & $\delta$($^\circ$) & $\alpha$(h)    &
                $n_{\mathrm{obs}}$ & $n_{b}$          &&
                $\Phi_{\nu}^{\mathrm{lim}}$           \\\hline
\multicolumn{7}{c}{ \emph{TeV Blazars} } & \multicolumn{7}{c}{ \emph{GeV Blazars} } \\
      Markarian 421  & 38.2 & 11.1 & 6 & 7.4 && 7.4&
      QSO 0219+428   & 42.9 &  2.4 & 5 & 5.5 && 9.6\\
      Markarian 501  & 39.8 & 16.9 & 8 & 6.4 && 14.7&
      QSO 0954+556   & 55.0 &  9.9 & 2 & 6.7 && 2.7\\
  \multicolumn{7}{c}{ \emph{Micro-quasars} } &  \multicolumn{7}{c}{ \emph{SNR \& Pulsars} }\\
      SS433          &  5.0 & 19.2 & 4 & 6.1 && 4.8&
      SGR 1900+14    &  9.3 & 19.1 & 5 & 5.7 && 7.8\\
      Cygnus X3      & 41.0 & 20.5 & 7 & 6.5 && 11.8&
      Crab Nebula    & 22.0 &  5.6 &10 & 6.7 && 17.8\\
      GRS 1915+105   & 10.9 & 19.3 & 7 & 6.1 && 11.2&
      Geminga        & 17.9 &  6.6 & 3 & 6.2 && 3.5\\
      Cygnus X1      & 35.2 & 20.0 & 8 & 7.0 && 13.2&
      Cassiopeia A   & 58.8 & 23.4 & 5 & 6.0 && 8.9\\
       \hline\hline
  \end{tabular}
 %\end{footnotesize}
  \caption[*]{
                $\nu_{\mu}+\nu_{\tau}$ flux limits from selected objects.
                $\delta$ is the declination in degrees, $\alpha$ the
		right ascension in hours,
                $n_{obs}$ is the number of observed events and $n_{b}$
                the expected
                background. $\Phi_{\nu}^{\mathrm{lim}}$ is the
                90\% CL upper limit on the flux of muon plus tau neutrinos in units of $10^{-8}\,\mathrm{GeV}^{-1}
                \mathrm{cm}^{-2}\mathrm{s}^{-1}$
                for a spectral index of 2
                and integrated above 10 GeV.}
 \end{center}
\end{table}

 The sensitivity for point sources can be increased by using a source-stacking
 analysis, where the data from the directions of sources known to have similar  
morphological characteristics are added. The background for each object is estimated 
off-source from the same zenith band as the location of the object, and added for all candidates.
A preliminary analysis performed with data collected in 2000-2003 for
 several types of sources shows no excess over the expected Poisson 
 statistics~\cite{AMANDA:stacking}.\\

\noindent{\bf b) Diffuse neutrino flux:} 
 Even if the neutrino flux from individual sources would be too weak to be detected with a detector 
of the size of AMANDA, a diffuse flux of neutrinos from the injected spectrum of all sources in the 
Universe could be detectable. The search for such flux is 
a challenge since it is, by definition, not correlated in time or 
position  with any particular object. The search is based 
on the expected harder neutrino spectrum, d$\Phi/dE_{\nu}\,\propto\,E_{\nu}^{-2}$, 
from the shock-acceleration of protons in the source, as
compared to the $E^{-3.7}$ dependence of the atmospheric neutrino flux. 
In analysis terms, this translates in exploiting the different shapes of the 
distribution of the number of optical modules hit, a variable related to the energy of the event. 
 This search can be done both for $\nu_{\mu}$ and cascades from $\nu_{e}$ and $\nu_{\tau}$, and can cover a wide 
range of energies, from a few tens of TeV to EeV. Above PeV energies, the Earth becomes 
opaque to neutrinos and the search has to be concentrated on events from near or above the horizon. \par
\begin{figure}[t]
\epsfxsize=5.0in
\centerline{\epsffile{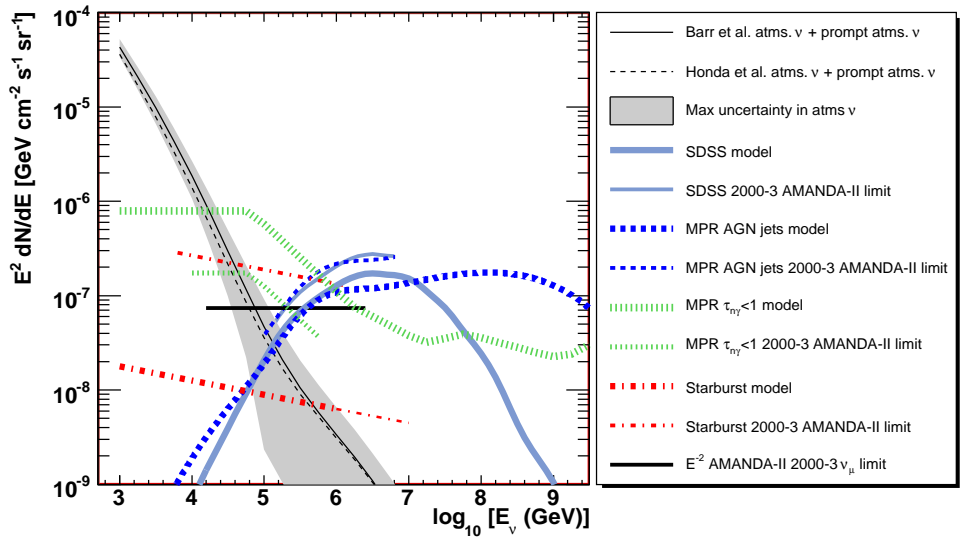}}
\caption[*]{AMANDA upper limits from the diffuse analysis. The Barr \textit{et al.}  and Honda
\textit{et al.} atmospheric neutrino models are shown as thin lines with
uncertainties represented by the band. Other models that were tested included the SDSS AGN core model
\cite{sdss}, the MPR upper bounds for AGN jets and optically
thin sources \cite{mpr}, and a starburst galaxy model~\cite{loeb_waxman_starburst}.
}
\label{fig:diffuse_limits}
\end{figure}
The most recent limit obtained by AMANDA for muon neutrinos is based on the 
analysis of four years of data (2000-2003), with a total lifetime of 807 
days~\cite{AMANDA:multiyear_diffuse}. The absence of a signal is translated to a 
limit on the diffuse muon neutrino flux of  $E_{\nu}^2\,d\Phi/dE_{\nu}\,<\,$7.4$\times 10^{-8}$ 
GeV cm$^{-2}$ s$^{-1}$ sr$^{-1}$ in the energy range 16 TeV--2.5 PeV. Other spectral shapes predicted by 
 different theoretical models~\cite{sdss, mpr,loeb_waxman_starburst} were also tested and limits set, 
as shown in figure~\ref{fig:diffuse_limits}. See~\cite{AMANDA:multiyear_diffuse} for details.\par
 A search for a diffuse flux in the cascade channel (sensitive to all flavours and 
with 4$\pi$ acceptance) is under way using 1000 days of live-time collected during 2000 to 2004.
Using 20\% of the data set a sensitivity on a $\nu_e$ flux of 2.7$\times 10^{-7}$ 
($E_{\nu}$/GeV)$^{-2}$ GeV cm$^{-2}$ s$^{-1}$ sr$^{-1}$ has been obtained~\cite{AMANDA:cascade_sensitivity}.\par
 In the ultra-high energy regime, $E_{\nu}\gtrapprox$ PeV, a search 
has been carried out on the 571 days of lifetime collected between 2000 and 2002  for a signal near the horizon. 
No statistically significant excess above the expected background has been seen, and 90\% CL upper limit on the diffuse 
all-flavor neutrino flux of $E_{\nu}^{2}\,d\Phi/dE_{\nu}\,<\, 2.7\;\times$ 10$^{-7}$ GeV cm$^{-2}$ s$^{-1}$
sr$^{-1}$ has been obtained, valid over the energy range of 2 $\times$ 10$^{5}$ GeV to 10$^{9}$ GeV~\cite{AMANDA:UHE}. 

\begin{figure}[t]
\begin{minipage}{0.48\linewidth}
\vspace*{-3.5cm}
\epsfxsize=\textwidth
\centerline{\epsffile{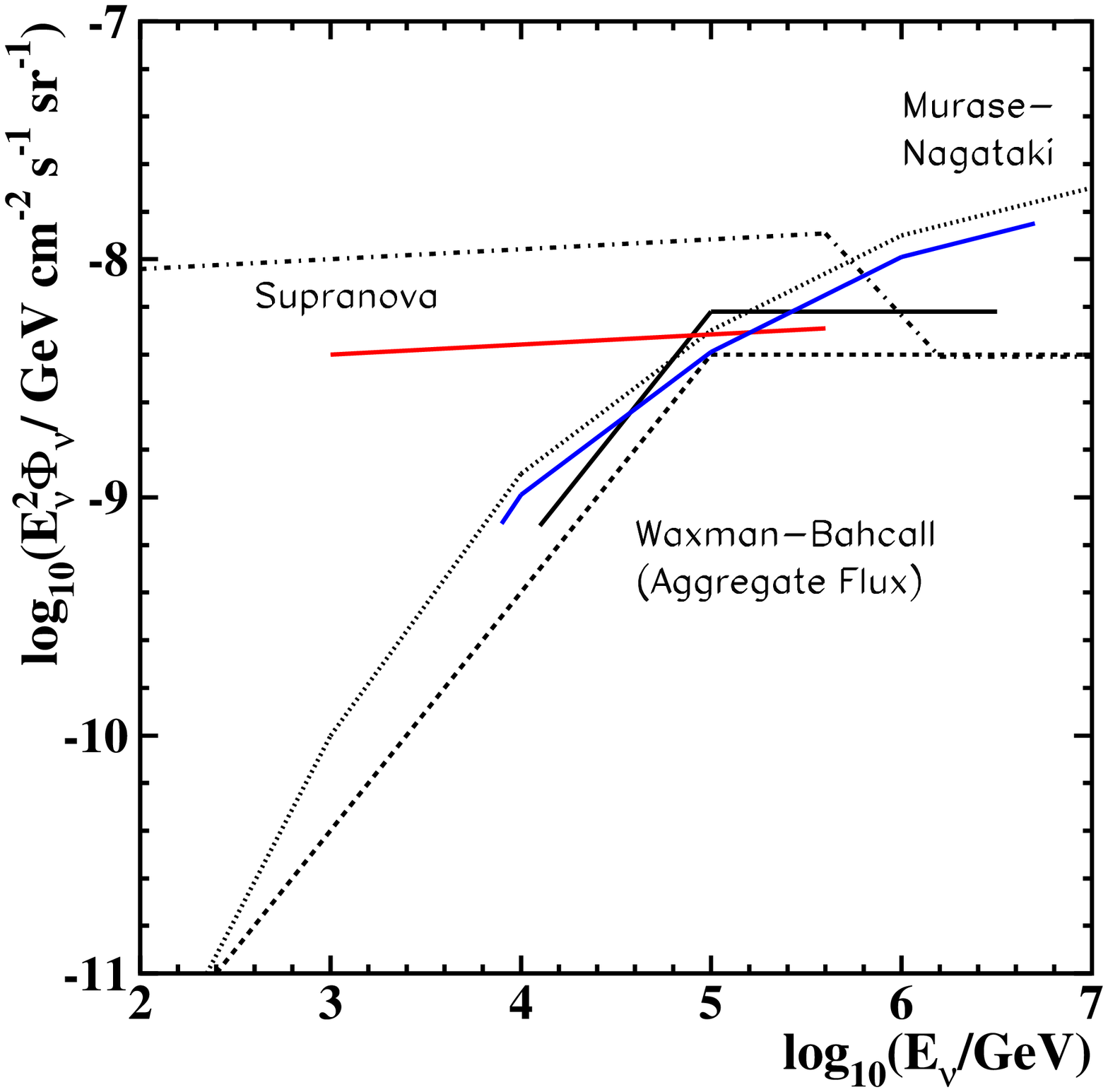}}
\caption[*]{
AMANDA GRB flux upper limits (solid lines) for $\nu_{\mu}$ energy spectra 
predicted by the Waxman-Bahcall spectrum (thick dashed line)~\cite{Waxman}, the Razzaque 
et al. spectrum (dot-dashed line)~\cite{Razzaq} and the Murase-Nagataki spectrum 
 (thin dotted line)~\cite{Murase}.)
}
\label{fig:grb_muon}
\end{minipage}
\hfill
\begin{minipage}[t]{0.48\linewidth}
\epsfxsize=1.1\textwidth
\centerline{\epsffile{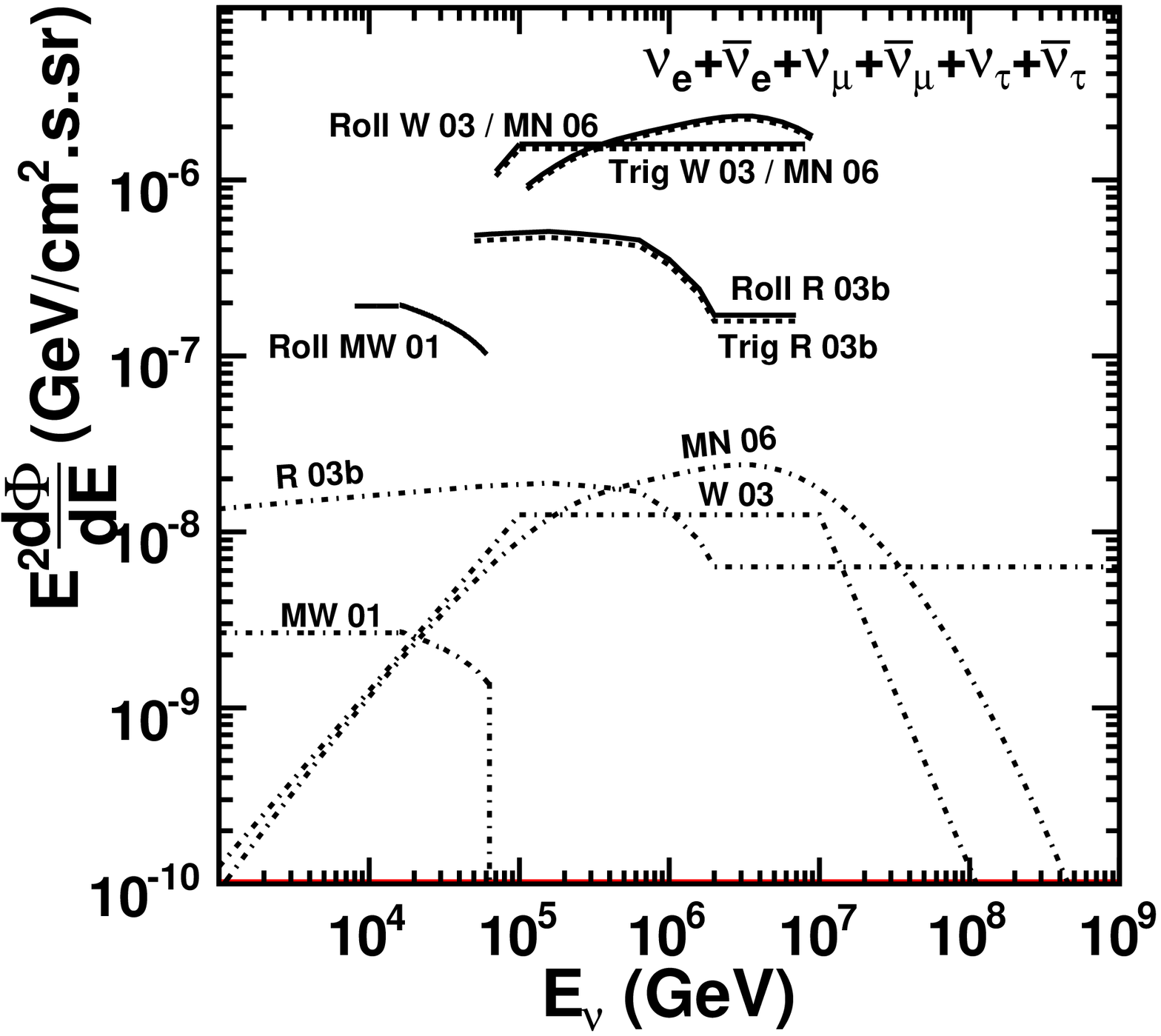}}
\caption[*]{
 Predicted GRB all-flavor neutrino fluxes compared to AMANDA cascade analysis limits (\textit{Rolling} limits (Roll) in solid line and
  \textit{Triggered} limits (Trig) in dashed line). 
Models shown are:~\cite{Waxman} (W~03),~\cite{supranova} (R~03b),~\cite{Murase} Model A (MN~06) and\cite{choke} (MW~01). 
  }
\label{fig:grb_casc}
\end{minipage}
\end{figure}

\noindent{\bf Gamma-ray bursts:} 
A rather special type of candidate neutrino point source is Gamma Ray Bursts (GRB), 
since for these objects one can have the time stamp and coordinates of the event from 
other detectors. This allows one to perform a practically background--free analysis using both 
off-source and off-time background estimation. An additional advantage of having the time stamp 
of the event is that the required pointing resolution can be relaxed and then
the cascade channel can be used, giving access to full--sky searches. 
Additionally, a ``rolling search'' can also be performed where no trigger information 
from any external detector is used and AMANDA data are searched for events clustered 
in short time periods.  
 We have performed three GRB search analyses, one for muon neutrinos~\cite{AMANDA:GRB_muon} and two using 
the cascade channel~\cite{AMANDA:GRB_cascade}. The triggered 
analyses rely on spatial and temporal correlations with photon observations 
of BATSE and several satellites of the Third Interplanetary Network (IPN).\par
 The muon neutrino analysis has been performed using 419 GRB bursts between 1997 and 2003. 
No neutrinos were observed in coincidence with the bursts, resulting in the most 
stringent upper limit on the muon neutrino flux from GRBs to date. Assuming a Waxman-Bahcall spectrum, 
a 90\% CL upper limit of $E_{\nu}^{2}\,d\Phi/dE_{\nu}\,{\leq}$ 6.0 $\times$ $10^{-9}\;\mathrm{GeV cm^{-2} s^{-1} sr^{-1}}$, 
has been obtained, with 90\% of the events expected within the energy range between 10 TeV and 3 PeV. 
We have also tested the flux predictions from several prominent GRB models based on averaged burst properties. 
The 90\% C.L. flux upper limits relative to these models are shown in figure~\ref{fig:grb_muon}.\par
 Concerning the cascade channel, we have performed two searches for neutrino-induced cascades. 
The triggered analysis searched for neutrinos in coincidence with 73 gamma-ray
bursts reported by BATSE in 2000. The rolling analysis searched for a
statistical excess of cascade-like events in time rolling windows of 1 s and 100 s in the period 2001 to 2003. 
The resulting limits are $E_{\nu}^{2}\,d\Phi/dE_{\nu}\,{\leq}$ 1.5 $\times 10^{-6}\; \mathrm{GeV cm^{-2} s^{-1} sr^{-1}}$ for the 
triggered analysis and $E_{\nu}^{2}\,d\Phi/dE_{\nu}\,{\leq}$ 1.6$\times 10^{-6}\; \mathrm{GeV cm^{-2} s^{-1} sr^{-1}}$ 
for the rolling analysis. Lacking spatial and temporal constraints, the rolling analysis has a reduced 
per-burst sensitivity relative to triggered analyses. On the other hand, a rolling analysis has the potential to detect 
sources missed by other methods. The test of specific models using the cascade channel is shown in 
figure~\ref{fig:grb_casc}.

\subsection{Search for dark matter candidates}\label{sec:DM}
 Searches for dark matter with neutrino telescopes are based on searches for an excess neutrino flux 
from relic particles gravitationally accumulated in the Earth or the Sun. 
 A massive (GeV-TeV range), weakly interacting and stable particle, the neutralino, appears in Minimally Supersymmetric 
extensions of the Standard Model that assume R-parity conservation, and it is a good candidate for non--baryonic dark matter. 
These relic particles, if accumulated in the center of the Sun or Earth, can annihilate pairwise, and neutrinos can result 
from the decays of the annihilation products~\cite{Jungman:96}. We have performed searches for neutralino dark matter accumulated in the 
Earth ($2001$\---$2003$ data set)~\cite{AMANDA:WIMP_earth}, and the Sun ($2001$ data set)~\cite{AMANDA:WIMP_sun}. The results are shown in figure~\ref{fig:wimp_limits}. 
The figures show the muon flux limit from neutralino annihilations, along  with the results from other indirect searches 
and predictions from theoretical models. Disfavoured models by recent direct searches~\cite{xenon10} are shown as (green/grey) dots.

\begin{figure}[t]
\begin{minipage}{0.48\linewidth}
\vspace*{-0.2cm}
\epsfxsize=\textwidth
\centerline{\epsffile{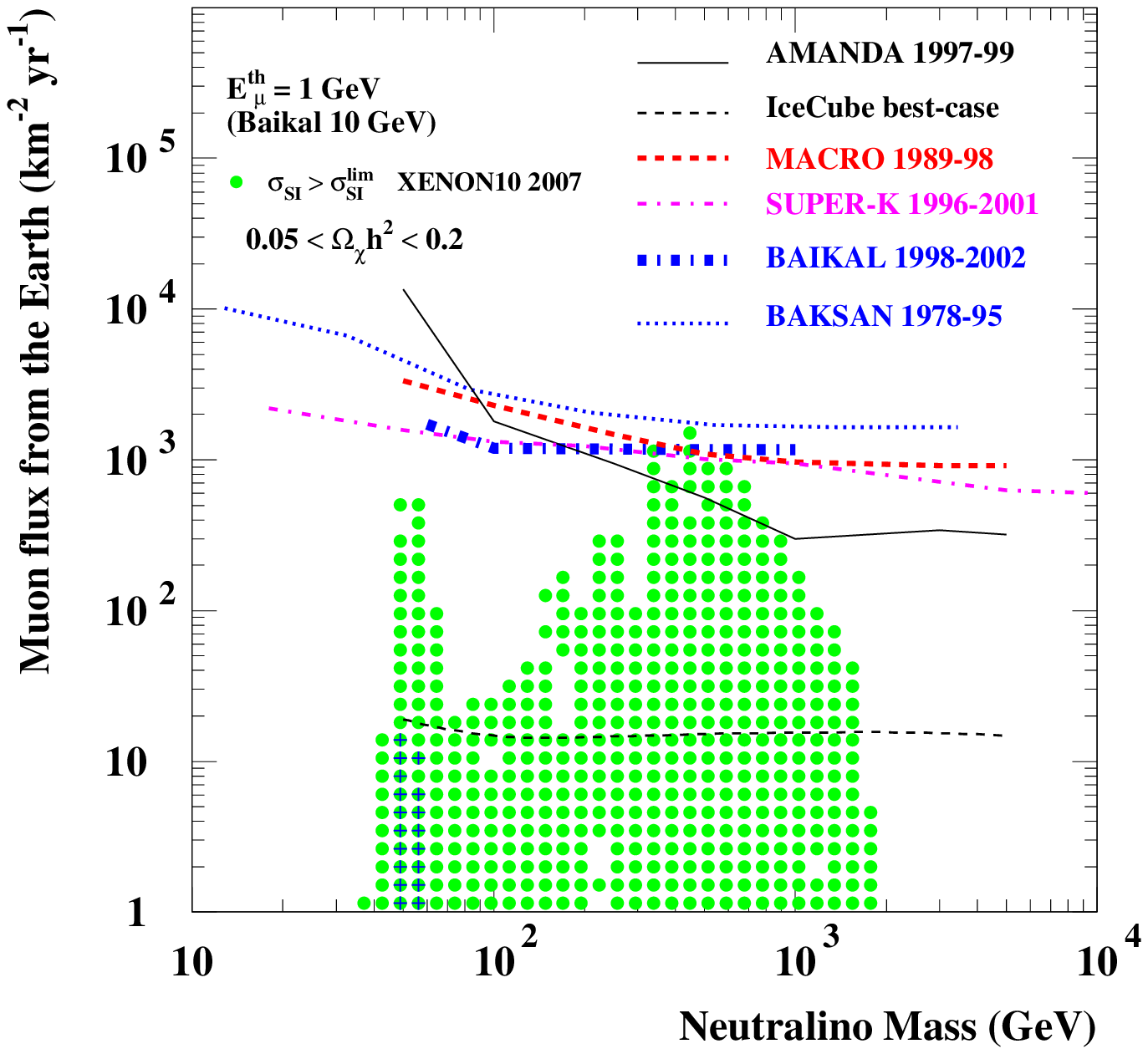}}
\end{minipage}
\hfill
\begin{minipage}{0.48\linewidth}
\vspace*{-0.2cm}
\epsfxsize=\textwidth
\centerline{\epsffile{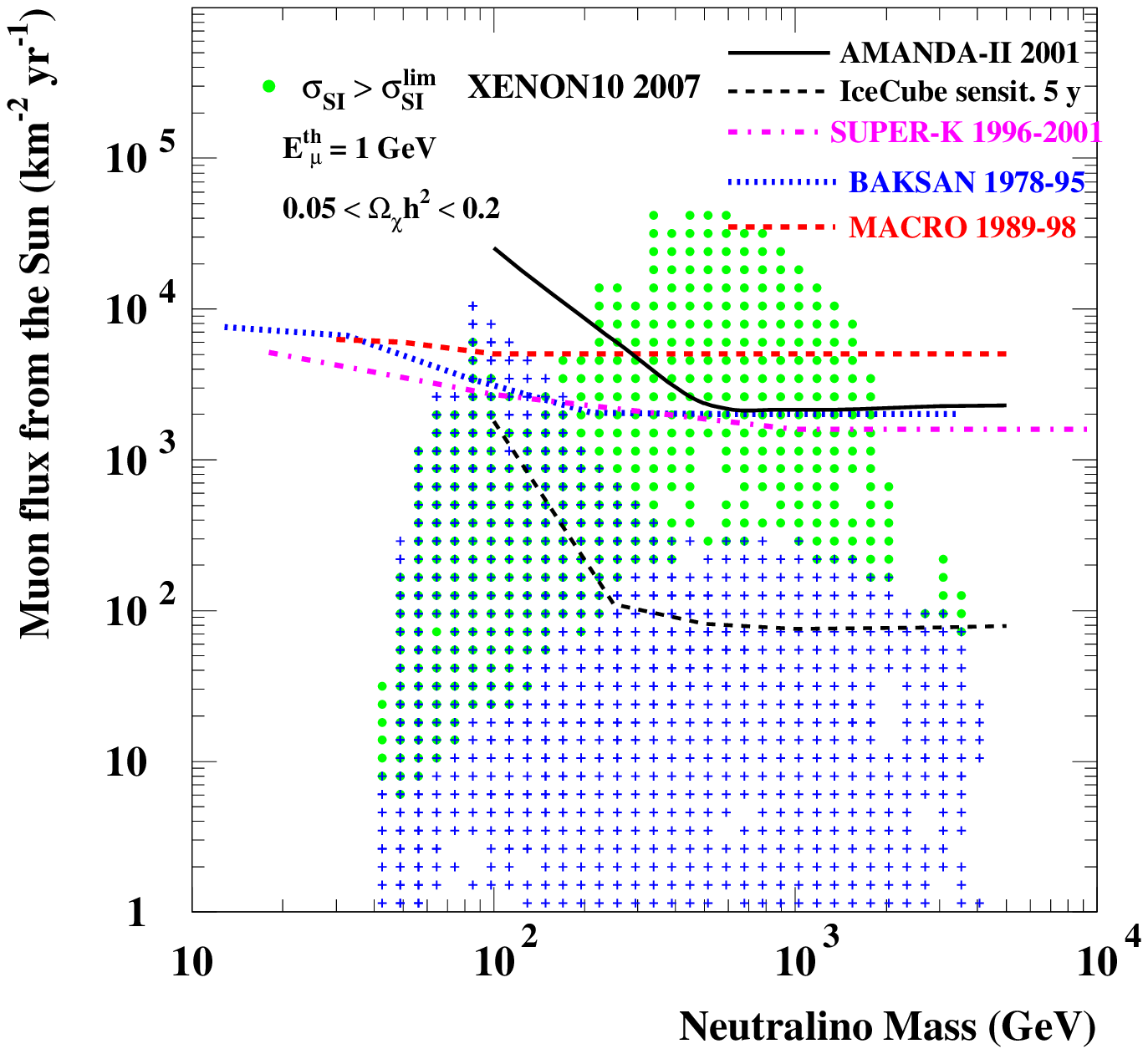}}
\end{minipage}
\caption[*]{$90\%$ CL upper limit on the muon flux from  neutralino annihilations in the center of the Earth (left) 
and from the Sun (right). Markers show predictions for cosmologically relevant MSSM models, the dots representing models 
excluded by XENON10~\cite{xenon10}.
}
\label{fig:wimp_limits}
\end{figure}

\subsection{Magnetic monopoles and exotics}\label{sec:monopoles}
 Stable magnetic monopoles with masses in the range between 10$^8$ to 10$^{17}$\,GeV are predicted in Grand Unified 
Theories~\cite{tHooft:1974}. They can be accelerated by large 
scale magnetic fields, and those with masses below $\sim10^{14}$\,GeV can acquire relativistic speeds. 
Their Cherenkov emission in ice is enhanced by a factor 8300 ($(n/2\alpha)^2$, where $n$ is the refractive index and $\alpha$ is the electromagnetic coupling constant), compared to a particle with unit electric charge 
and the same speed. This constitutes the main experimental signature of monopoles in neutrino telescopes: extremely 
bright events.  Monopoles with masses above $10^{11}$\,GeV can cross the entire Earth
and enter the detector from below, allowing for a search of 
up-going particles, practically background free.\par
 We have analyzed data taken with  AMANDA during the year 2000 in search for monopole candidates~\cite{AMANDA:monopole}. 
For monopole speeds greater than $\beta=0.8$ and masses greater than $\sim10^{11}$\,GeV,
the flux limit is presently the most stringent experimental limit. Figure \ref{fig:monopole} shows the flux 
limits set by AMANDA compared to those set by MACRO~\cite{MACRO:monopole} and by BAIKAL~\cite{BAIKAL:monopole}.\par

\begin{figure}[t]
\begin{minipage}{0.48\linewidth}
\epsfxsize=0.9\textwidth
\epsfysize=0.8\textwidth
\centerline{\epsffile{plots/Monopole_sysLimitsColor02.epsi}}
\caption[*]{Preliminary limits on the flux of relativistic magnetic monopoles set by AMANDA. 
Earlier AMANDA (marked AMANDA-B10), MACRO and BAIKAL results are also shown.}
\label{fig:monopole}
\end{minipage}
\hfill
\begin{minipage}{0.48\linewidth}
\vspace{-0.9cm}
\epsfxsize=3.0in
\centerline{\epsffile{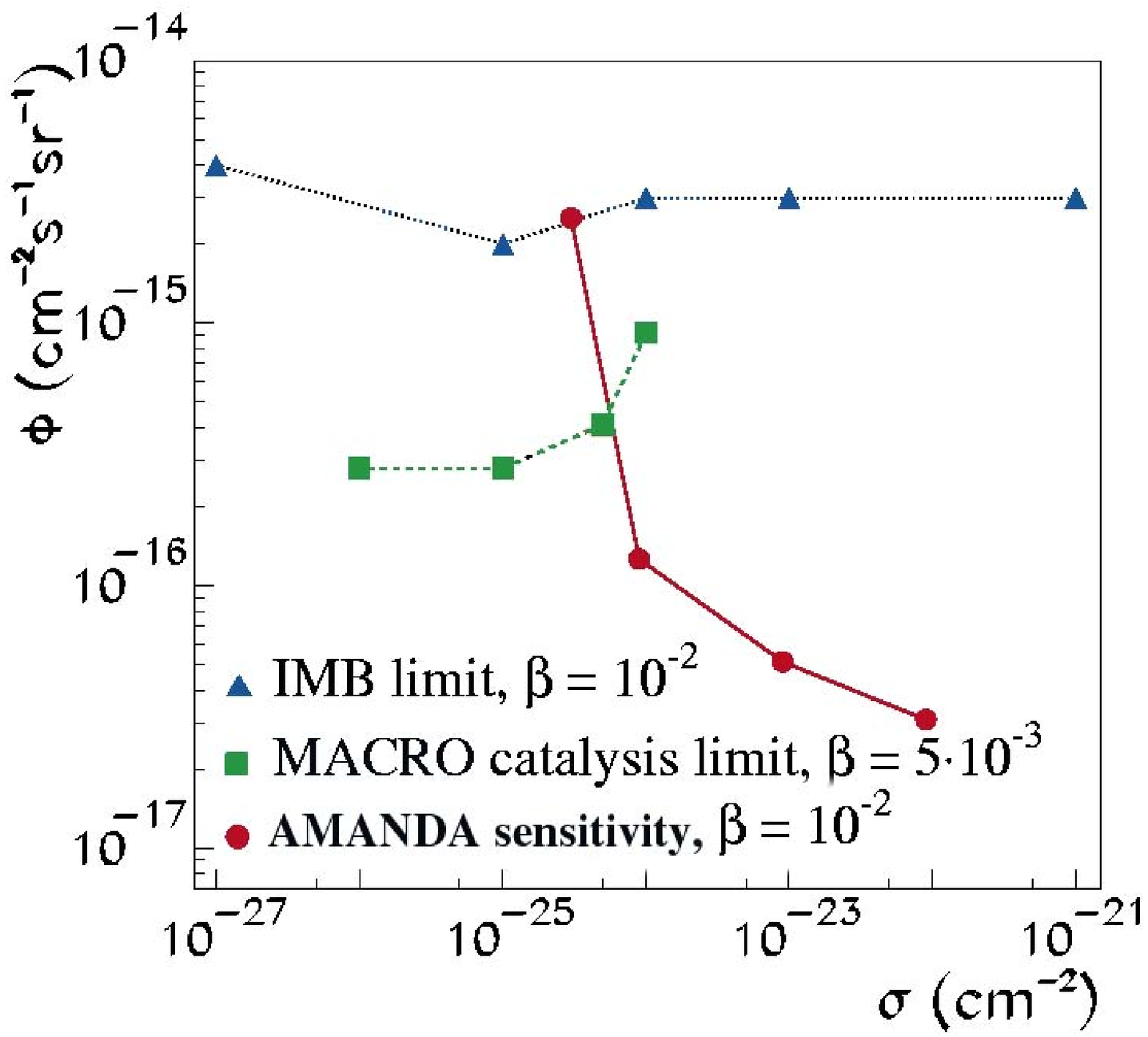}}
\caption[*]{Preliminary AMANDA sensitivity to sub-relativistic particles as a function of varying catalysis cross section. 
Limits set by IMB and MACRO are also shown.
}
\label{fig:subrelativistic}
\end{minipage}
\end{figure}

Supermassive monopoles will remain sub-relativistic and do not 
produce Cherenkov radiation when passing through ice. However, 
they can be detected through  nucleon decay catalysis: the charged decay products 
($e$'s, $\pi$'s, $\mu$'s and $K$'s) will emit Cherenkov radiation along the monopole trajectory. 
A similar process applies for neutral Q-balls, another type of relic massive particles~\cite{kusenko98b}. 
  In a neutrino telescope, the signature of catalyzing particles would be 
a series of closely spaced electromagnetic showers produced along the particle trajectory. 
The detection of slow particles builds on the fact that relativistic muons emit light during $\sim\!3\,\mu$s 
(the time it takes to cross the AMANDA volume), whereas slow particles would emit during a large fraction of 
the 33\,$\mu$s time window of the AMANDA data acquisition system. 
 The sensitivity of a preliminary search for sub-relativistic particles in 113 days of lifetime in 
2001~\cite{AMANDA:subrelativistic} is shown in figure~\ref{fig:subrelativistic}, along with existing limits 
from other experiments~\cite{MACRO:subrelativistic,becker94}.

\subsection{Search for non-standard neutrino oscillations}\label{sec:nuosc}
 Some phenomenological models of physics beyond the Standard Model predict 
flavour mixing in the neutrino sector in addition to the standard mass-induced 
oscillations~\cite{Coleman:97a}. In particular, violation of Lorentz Invariance can lead to 
different maximum attainable velocities (MAV) for the different flavours, 
and therefore to MAV-induced oscillations, since MAV eigenstates will not be 
flavour eigenstates. The effect can be parametrized in terms of $\delta$c/c, 
the difference in maximal attainable velocity. 
In contrast to mass-induced oscillations, MAV oscillations show a linear 
energy dependence of the oscillation frequency. The expected signature in 
AMANDA/IceCube is a distortion of the angular and energy spectra of 
atmospheric neutrinos at energies above 10$^5$\,GeV. \par
We have used data collected between 2000 and 2003 to search for anomalous 
oscillation effects in 3401 atmospheric neutrinos collected in that 
period~\cite{AMANDA:non_standard_oscillations}.  
The exclusion regions of $\delta$c/c as a function of the mixing angle $\Theta_c$ at different 
confident levels are shown in figure~\ref{fig:mva_limits}, for a particular 
value of the unconstrained phase $\eta$. However the results can be shown to be quite 
insensitive to the value of $\eta$.
\begin{figure}[t]
\begin{minipage}{0.48\linewidth}
\vspace{-0.5cm}
\epsfxsize=1.05\linewidth
\centerline{\epsffile{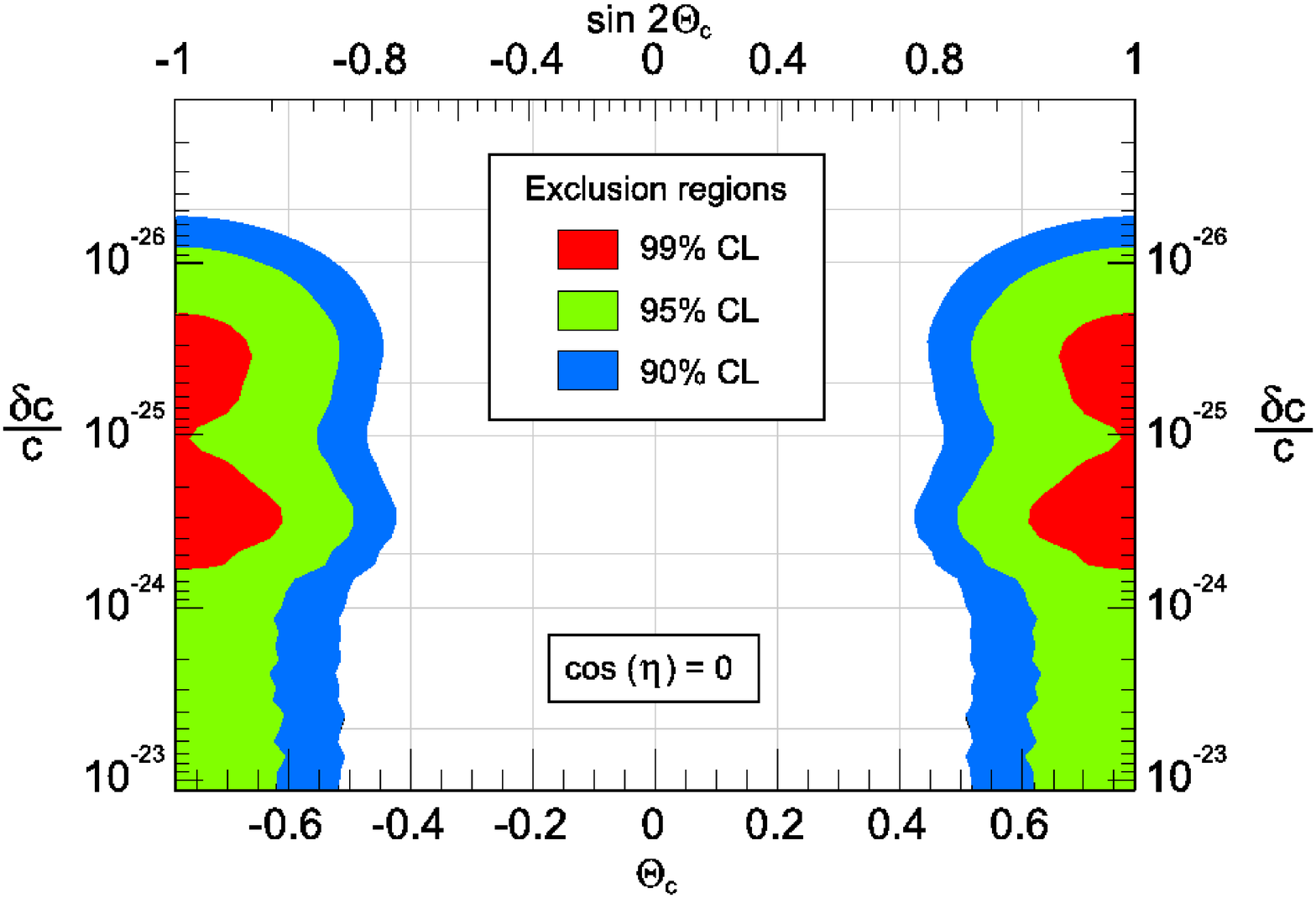}}
\caption[*]{Preliminary exclusion regions for $\delta$c/c as a function of the mixing angle $\Theta$}
\label{fig:mva_limits}
\end{minipage}
\hfill
\begin{minipage}{0.48\linewidth}
%\vspace{-0.5cm}
%\epsfxsize=3.0in
\epsfxsize=\linewidth
\centerline{\epsffile{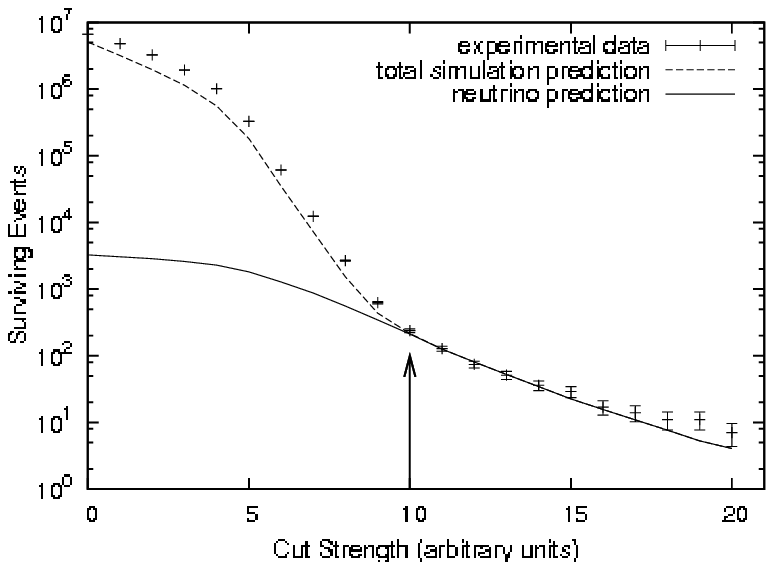}}
\caption[*]{Number of events remaining for data, atmospheric muons and atmospheric neutrinos as a function of 
increasing event quality criteria.}
\label{fig:icecube_cuts}
\end{minipage}
\end{figure}

\section{First results and perspectives from IceCube}
\label{sec:IC_results}

 IceCube is growing rapidly. The detector took data with 9 strings during 2006 
and, after the deployment season during the austral summer 2006/07, it consists of 22 strings and 
a total of 1320 DOMs. In addition 26 IceTop tanks are operational. 
With 98\% of the DOMs deployed so far commissioned and working, the detector parameters meet 
or exceed the design specifications. Construction is on schedule and foreseen to be completed by 2011.\par
The data from 137 days of lifetime accumulated during 2006 has 
been analyzed and the first atmospheric neutrino 
 candidates identified, a total of 234 for an expected yield of 211 $\pm \,76(syst.) \pm\,14(stat.)$ from 
a pure atmospheric neutrino flux~\cite{IceCube:atmospheric}. Figure~\ref{fig:icecube_cuts} shows the number 
of data events, atmospheric muon background and atmospheric neutrinos remaining as 
a function of increasing event quality selection. Figure~\ref{fig:icecube_zenith} shows the zenith 
angle distribution of the 234 events. Some residual atmospheric muons remain near the horizon, but 
the sample is consistent with the atmospheric neutrino angular distribution (shown as a grey band 
including theoretical uncertainties in the expected flux and systematic uncertainties in the detector response)
 for declinations above $\sim$120$^{\circ}$. 
The data was also searched for ``hot spots'' due to point sources~\cite{IceCube:point}, and a preliminary significance sky map 
obtained, as shown in~\ref{fig:icecube_skyplot}. 
The resulting preliminary sky-averaged point-source sensitivity for an $E^{-2}$ spectrum is $E_{\nu}^{2}\,d\Phi/dE_{\nu} = 12\times 10^{-8}\ \mathrm{GeV cm^{-2}\ s^{-1}}$, already comparable with the results of 
5 years of AMANDA data (see table~1.1).\par
 The sensitivity of IceCube to the different physics topics that can be addressed will increase 
rapidly as exposure and size increase during construction. The expected sensitivity to a diffuse flux 
of the complete detector after one year of data taking is  
 $E_{\nu}^{2}\,d\Phi/dE_{\nu}\,\sim 8.0\,\times 10^{-9}$ $\mathrm{GeV cm^{-2} s^{-1} sr^{-1}}$, 
an order of magnitude below the current AMANDA limit (obtained with three years of exposure). A similar 
improvement is expected in the sensitivity to point sources, not only due to the bigger effective 
volume but also due to the expected sub-degree angular resolution at TeV energies. \par
The possibility to identify flavour is one of the significant improvements of the capabilities of 
IceCube with respect to AMANDA. IceCube will be able to identify the typical ``double-bang'' signature  
of $\nu_{\tau}$  events, the hadronic shower at the interaction vertex and the $\tau$ decay 
shower. Above PeV energies a tau can travel $(O)$100 m, and the separation of the two showers 
can be resolved. AMANDA is too small for this. 
 
\begin{figure}[t]
\begin{minipage}{0.48\linewidth}
%\vspace*{0.5cm}
\epsfxsize=1.\linewidth
\centerline{\epsffile{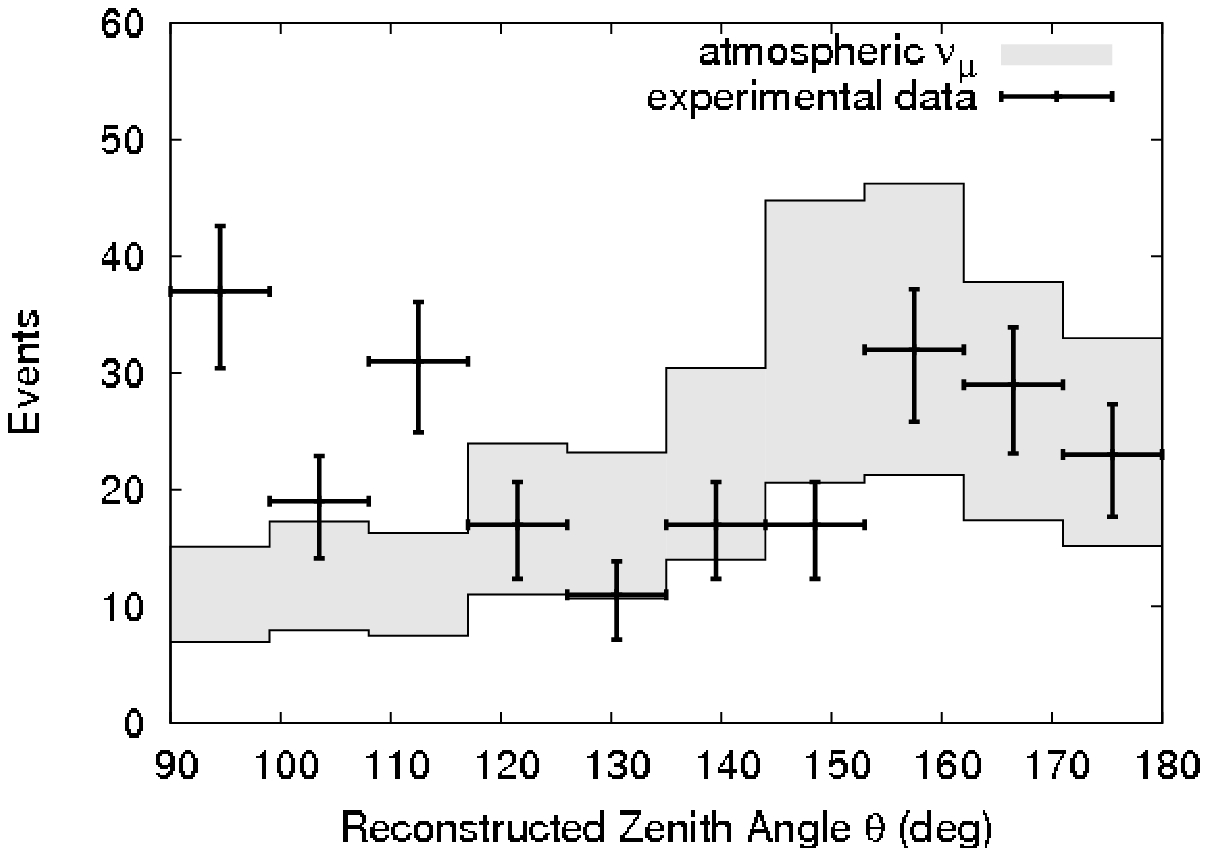}}
\caption[*]{Zenith angular distribution of the 234 neutrino candidates in the IceCube 2006 data set.}
\label{fig:icecube_zenith}
\end{minipage}
\hfill
\begin{minipage}{0.48\linewidth}
%\vspace*{-1.cm}
\epsfxsize=1.05\linewidth
\epsfysize=0.65\linewidth
\centerline{\epsffile{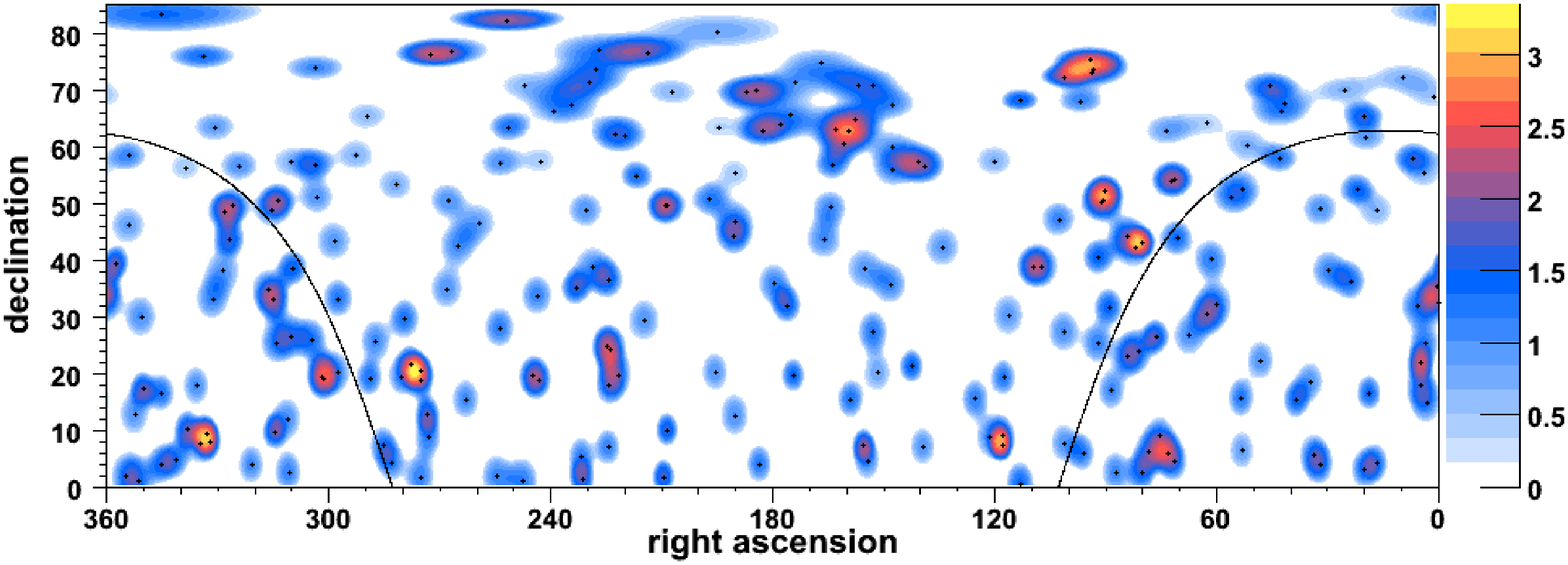}}
\caption[*]{Significance sky map using 137 days lifetime of IceCube data in 2006. 
The figure shows the deviation from a uniform background.}
\label{fig:icecube_skyplot}
\end{minipage}
\end{figure}

\section{Future extensions}\label{sec:future}
\vspace*{-0.3cm}
There are ongoing plans to extend the capabilities of IceCube both at lower energies and 
at higher energies.  On the low energy side we will build a compact core of 6 strings in the middle of 
the IceCube array, with typical inter-string separation of $\sim$50~m. This allows us to use the 
surrounding strings as a veto region in order to define contained events and reduce 
the atmospheric muon background. Such a compact core will make it possible to increase the sensitivity 
to events below 100~GeV, an important energy range for the dark matter searches. The ability 
to reduce the background in such way will allows us to ``look'' to the Sun continuously, even when 
above the horizon. This also opens the possibility of looking at the Galactic center. \par
 Above neutrino energies of $\sim$10$^{18}$ eV the radio and acoustic signal of the particle 
cascade produced at the neutrino-nucleon interaction point dominates over the optical 
Cherenkov emission. Ice is extremely transparent to both radio and acoustic signals  
in the frequency ranges of interest (MHz-GHz for radio, kHz for acoustic), with attenuation 
lengths of the order of km in both cases. Two exploratory projects, AURA (Askaryan Under ice Radio Array)~\cite{AURA} 
and SPATS (South Pole Acoustic Test Setup)~\cite{SPATS} are being carried out to explore the characteristics 
of polar ice and to develop and assess different hardware options for a future hybrid array of 90 strings with 
1 km spacing surrounding the current IceCube site.

\normalsize
\section*{\normalsize \bf References} 

* see www.icecube.wisc.edu/collaboration/authorlists/2007/5.html for full author list and acknowledgments.


\begin{thebibliography}{99}

\bibitem{Hess:36a} V. Hess, Nobel Lecture, in {\it Nobel Lectures, Physics 1922-1944}. Elsevier Publishing, Amsterdam (1965).

\bibitem{IceCube:Performance} A. Achterberg {\it et al}., Astropart. Phys. {\bfseries 26}, 155, (2006).

%%  Galactic plane ICRC05
\bibitem{AMANDA:galactic_plane} J.~L. Kelley, Proc. of ICRC05, Pune, India, {\bfseries 00}, 101, (2005). 

%% 5yr point source
\bibitem{AMANDA:5yr_pointsource}  A. Achterberg {\it et al}., Phys. Rev. {\bfseries D75}, 102001 (2007).

%%  Stacking
\bibitem{AMANDA:stacking} A. Gross {\it et al.}, in proceedings of ICRC05, Pune, India, (2005).

%%  multiyear diffuse
\bibitem{sdss} F.~W. Stecker {\it et al}., Phys. Rev. Lett. \textbf{66}, 2697 (1991); \textbf{69}, 2738(E) (1992);\textbf{72}, 107301, (2005).
\bibitem{mpr} K. Mannheim, R.~J. Protheroe and J.~P. Rachen, Phys. Rev. D \textbf{63}, 023003, (2000).
\bibitem{loeb_waxman_starburst} A. Loeb and E. Waxman, J. Cosmol. Astropart. Phys. JCAP05, 003 (2006).
\bibitem{AMANDA:multiyear_diffuse}  A. Achterberg {\it et al}., Phys. Rev. {\bfseries D76}, 042008 (2007).

%%  Cascade
\bibitem{AMANDA:cascade_sensitivity} O. Tarasova, M. Kowalski and M. Walter, Proc. of ICRC07, M\'erida, Mexico, (2007).
arXiv:0711.0353, page 83.
%%  UHE
\bibitem{AMANDA:UHE} M. Ackermann {\it et al}., to appear in  Astrophys. Journal. arXiv:0711.3022.

%%  GRB muon
\bibitem{AMANDA:GRB_muon} A. Achterberg {\it et al}., to appear in  Astrophys. Journal. arXiv:0705.1186.

%%  GRB Cascade
\bibitem{AMANDA:GRB_cascade} A. Achterberg {\it et al}., Astrophys. Journal. {\bfseries  664}, 397, (2007).

\bibitem{Waxman} E. Waxman, Nucl. Phys. B Proc. Supp. {\bfseries 118},  353, (2003).
\bibitem{Razzaq} S. Razzaque {\it et al}.,  Phys. Rev. {\bfseries D68}, 3001, (2003).
\bibitem{Murase} K. Murase and S. Nagataki, Phys. Rev. {\bfseries D73}, 063002, (2006). 
\bibitem{supranova} S.~Razzaque, P.~Mesz\'aros and E.~Waxman,  Phys. Rev. Lett. {\bfseries 90}, 241103,  (2003).
\bibitem{choke} P. Mezs\'aros and E. Waxman, Phys. Rev. Lett. {\bfseries 87}, 171102, (2001).

%% WIMPs
\bibitem{Jungman:96} G. Jungman, M. Kamionkowski and K. Griest, Phys. Rep. {\bfseries 267}, 195, (1996).
\bibitem{AMANDA:WIMP_earth} A. Achterberg  {\it et al}., Astropart. Phys. {\bfseries 26}, 129, (2006).
\bibitem{AMANDA:WIMP_sun} M. Ackermann {\it et al}.,  Astropart. Phys. {\bfseries 24}, 459, (2006).
\bibitem{xenon10} J. Angle {\it et al}., arXiv:0706.003, (2007).
%\bibitem{indirect} M. Boliev {\it et al}.,  in \textit{Proc. of Dark Matter in Astro- and Particle Physics}, edited by H.V. Klapdor-Kleingrothaus and Y. Ramachers (World Scientific, 1997); M. Ambrosio {\it et al}., Phys. Rev. \textbf{D60}, 082002, (1999); S. Desai {\it et al}., Phys. Rev. \textbf{D70}, 083523, (2004), erratum \textit{ibid} \textbf{D70}, 109901, (2004); V. Aynutdinov {\it et al}., in \textit{Proc. of First Workshop on Exotic Physics with Neutrino Telescopes (EPNT06)}, edited by C. de los Heros (Uppsala University, 2006) (arXiv:astro-ph/0701333),

%% monopoles
\bibitem{tHooft:1974} G.~'t Hooft, Nucl. Phys. B\textbf{79}, 276, (1974).
\bibitem{AMANDA:monopole} H. Wissing, Proc. of ICRC07, M\'erida, Mexico, (2007). arXiv:0711.0353, page 139.
\bibitem{MACRO:monopole}  M. Ambrosio {\it et al}., Eur. Phys. J. {\bfseries C25}, 511, (2002).
\bibitem{BAIKAL:monopole} The BAIKAL Collaboration,  Proc. of ICRC05, Pune, India (2005).

%% subrelativistic
%% \bibitem{derujula84}  A.~De R\'ujula and S.~L.~Glashow, Nature \textbf{312},  734, (1984).
\bibitem{kusenko98b}  A.~Kusenko  {\it et al}., Phys. Rev. Lett. \textbf{80},  3185, (1998).
\bibitem{MACRO:subrelativistic}  M. Ambrosio {\it et al}., Eur. Phys. J. {\bfseries C26}, 163, (2002).
\bibitem{becker94}  R.~Becker-Szendy {\it et al}.,  Phys. Rev.  {\bfseries D49}, 2169, (1994).
\bibitem{AMANDA:subrelativistic} A. Pohl and D. Hardtke, Proc. of ICRC07, M\'erida, Mexico, (2007). 
arXiv:0711.0353, page 143.

%% oscillations
\bibitem{Coleman:97a} S.~Coleman and S.~L.~Glashow, Phys. Lett {\bfseries B405}, 249, (1997).
\bibitem{AMANDA:non_standard_oscillations} J.~Ahrens and J.~L.~Kelley, Proc. of ICRC07, M\'erida, Mexico, (2007). 
arXiv:0711.0353, page 55.

%IceCube9 atmospheric
\bibitem{IceCube:atmospheric}A. Achterberg  {\it et al}., Phys. Rev. {\bfseries D76}, 027101, (2007).

%IceCube9 point source
\bibitem{IceCube:point} C. Finnley, J. Dumm and T. Montaruli, Proc. ICRC07, M\'erida, Mexico, (2007). 
arXiv:0711.0353, page 107.

%%AURA
\bibitem{AURA} H. Landsman, L. Ruckman and  G.~S.~Varner, Proc. of ICRC07, M\'erida, Mexico, (2007). 
arXiv:0711.0353, page 163.
%%SPATS
\bibitem{SPATS} S. B\"oser {\it et al},  Proc. of ICRC07, M\'erida, Mexico, (2007)
\end{thebibliography}
\end{document}